\documentclass[11pt,a4paper,english,superscriptaddress]{revtex4}

\usepackage{amssymb}
\usepackage{amsmath}
\usepackage{graphicx}
\usepackage{babel}
\usepackage{hyperref}
\usepackage{color}

\begin{document}

\title{Renormalization group improvement of the effective potential in six dimensions}

\author{Huan Souza}
\email{huan.souza@icen.ufpa.br}
\affiliation{Faculdade de F\'isica, Universidade Federal do Par\'a, 66075-110, Bel\'em, Par\'a, Brazil.}

\author{L.~Ibiapina~Bevilaqua}
\email{leandro@ect.ufrn.br}
\affiliation{Escola de Ci\^encias e Tecnologia, Universidade Federal do Rio Grande do Norte\\
Caixa Postal 1524, 59072-970, Natal, Rio Grande do Norte, Brazil}

\author{A. C. Lehum}
\email{lehum@ufpa.br}
\affiliation{Faculdade de F\'isica, Universidade Federal do Par\'a, 66075-110, Bel\'em, Par\'a, Brazil.}


\begin{abstract}
 Using the renormalization group improvement technique, we study the effective potential of a model consisting of $N$ scalar fields $\phi^i$ transforming in the fundamental representation of $O(N)$ group coupled to an additional scalar field $\sigma$ via cubic interactions, defined in a six-dimensional spacetime. We find that the model presents a metastable vacuum, that can be long-lived, where the particles become massive. The existence of attractive and repulsive interactions plays a crucial role in such phenomena.
\end{abstract}

\maketitle

\section{Introduction}

Toy models have been intensely explored in scientific literature, since they provide good theoretical
laboratories to discuss key concepts of quantum field theory. Although we might have an unrealistic theory, it may highlight some interesting features we want to study. 

An instance of such toy models is the theory of a scalar field with cubic interaction in six dimensions. The $\phi^3_6$ model has been used to discuss a wide variety of topics. For example, this
model shares with QCD the interesting phenomenon of asymptotic freedom \cite{Srednicki:2007}, but is considerably simpler than the latter, thus providing a useful tool to explore this phenomenon
\cite{Gracey:2020baa}. Unlike QCD, however, this model has an unbounded potential from below and, although we might arrange for a stable local minimum, this stability is lost at a critical temperature
\cite{Altherr:1991fu}. This model was also used to study the behaviour of quantum gravity models with thermal instability \cite{Brandt:2008cn}. Moreover, some variations of this model are also fruitful in ideas. 
In \cite{Grosse:2006tc}, for example, the authors quantized and solved the noncommutative $\phi^3_6$ and were also able to compute the exact renormalization of the wave-function and coupling constant by mapping it to the Kontesevich model. 

In more recent years, the interest in a particular model with $N+1$ scalar fields in $d=6-\epsilon$ coupled via cubic interactions has grown \cite{Fei,Fei:2014xta,Gracey:2015tta,Arias-Tamargo:2020fow,Eichhorn:2016hdi}. 
This model is described by the Lagrangian
\begin{eqnarray}\label{o(n)-lagrangian}
 \mathcal{L}=\frac{1}{2}(\partial_\mu\phi^i)^2+\frac{1}{2}(\partial_\mu\sigma)^2+\frac{g_1}{2}(\sigma\phi^i\phi^i)+\frac{g_2}{6}\sigma^3, \qquad (i=1,2,\cdots, N),
\end{eqnarray}
and it was argued in \cite{Fei} that it provides an UV completion to the $O(N)$ symmetric scalar field theory with interaction $(\phi^i\phi^i)^2$ in the dimension range $4<d<6$, at least for large $N$.

As it is well known, spontaneous symmetry breaking is one of such key concepts in particle physics, with Higgs mechanism playing a fundamental role in the Standard Model. 
In that case, the symmetry breaking requires a mass parameter in the Lagrangian but S. Coleman and E. Weinberg demonstrated in \cite{Coleman:1973jx} that a spontaneous symmetry breaking may occur due to 
radiative corrections when a quadratic mass term is absent from the Lagrangian, as it is the case in conformally-invariant theories, such as the $\phi^3_6$ model, where we have a dimensionless coupling constant.

In order to discuss the Coleman-Weinberg (CW) mechanism, the standard procedure is to compute the effective potential, a powerful and convenient tool to explore many aspects of the the low-energy 
sector of a quantum field theory. In several situations, the one-loop approximation is good enough, but of course we want sometimes to improve it, adding higher-order contributions in the loop 
expansion. However, since calculations become very complicated already at two-loop, some techniques were developed to improve the calculation of the effective potential. In particular, we cite 
\cite{McKeon:1998tr}, where the effective action for the $\phi^3_6$ model was explicitly computed observing that the appearance of an arbitrary mass scale $\mu^2$ introduced by renormalization imposes some conditions for the quantum corrections to the classical potential. The so-called renormalization 
group improvement has been intensely used to go beyond one-loop approximation \cite{Ahmady:2002qg,Chishtie:2006ck,Elias:2003zm,Chishtie:2005hr,Elias:2004bc,Lehum:2019msl, Meissner:2008uw, Dias:2014txa, Quinto:2014zaa}. The general idea is to use the renormalization group equations (RGE) to sum up sub-series of the effective potential.

In this work, we compute the improved effective potential and use it to discuss the vacuum structure of a massless theory of scalars with cubic interaction in six dimensions. Our model consists of $N$ 
scalar fields $\phi^i$ transforming in the fundamental representation of $O(N)$ coupled to an additional scalar field $\sigma$ via cubic interactions, described by the Lagrangian 
(\ref{o(n)-lagrangian}). This theory has a potential unbounded from below, but it is nevertheless possible that radiative corrections might generate a stable false vacuum \cite{Gonzalez:2017hih}. Our results indicate that the CW mechanism does indeed provide a metastable vacuum and a generation of mass.

This work is organized as follows, in section II we compute the effective potential using the Renormalization Group Equation and explore some of its properties in $d=6$ dimensions. In section III, we draw our conclusions.
 
\section{The effective potential in $d=6$ dimensions}
 
We start by using the RGE to evaluate the effective potential for the model defined by the Lagrangian (\ref{o(n)-lagrangian}) in $d=6$ dimensions. The effective potential will be computed to the $\sigma$ 
field, including quantum fluctuations due to $\phi_i$ and $\sigma$ interactions, but we are assuming that $\langle\phi_i\rangle=0$ (so the $O(N)$ symmetry of this sector of the theory is kept manifest). 
That means $\sigma$ is the only degree of freedom in the effective potential. This choice is enough to study a possible generation of mass in such theory, as we discuss in the appendix.  

Following the prescription for the RG improvement technique \cite{McKeon:1998tr}, we start assuming that the effective potential has to satisfy the RGE:
\begin{equation}\label{rge1}
 \Bigr(\mu\frac{\partial}{\partial\mu}+ \beta_{g_1}\frac{\partial}{\partial g_1} + \beta_{g_2}\frac{\partial}{\partial g_2} +\gamma_\sigma\frac{\partial}{\partial\sigma}\Bigr)V_{eff}(\sigma)=0,
\end{equation}
\noindent where $\beta_{g_1}$ and $\beta_{g_2}$ are the two $\beta$-functions to this model and $\gamma_{\sigma}$ is the anomalous dimension for the scalar field $\sigma$.

In order to determine the effective potential, it is useful to write $V_{eff}$ as
\begin{equation}\label{ansatz1}
 V_{eff}=\frac{1}{6}\sigma^3S_{eff}\big(g_1,g_2,L(\sigma)\big),
\end{equation}

\noindent where $S_{eff}(g_1,g_2,L(\sigma))$ is a function of the coupling constants and $L(\sigma)=\ln\frac{\sigma^2}{\mu^2}$.

Now we observe that
\begin{eqnarray}
 \mu\frac{\partial V_{eff}}{\partial\mu} & =& -2\frac{\sigma^3}{6}\frac{\partial S_{eff}}{\partial L}=-2\frac{\partial V_{eff}}{\partial L} \\
 \sigma\frac{\partial V_{eff}}{\partial\sigma} & = &\frac{1}{6}\sigma^3\left(3+2\frac{\partial}{\partial L}\right)S_{eff}=\left(3+2\frac{\partial}{\partial L}\right)V_{eff},
\end{eqnarray}
so we can rewrite (\ref{rge1}) in terms of derivatives with respect to $L$ and thus we find the RGE for $S_{eff}$ to be
\begin{eqnarray}\label{rge_seff}
\Big[2(-1+\gamma_\sigma)\frac{\partial}{\partial L} + \beta_{g_1}\frac{\partial}{\partial g_1} + \beta_{g_2}\frac{\partial}{\partial g_2} + 3\gamma_\sigma\Big]S_{eff}=0.
\end{eqnarray} 

The one-loop renormalization group functions for the model (\ref{o(n)-lagrangian}) were computed in \cite{Fei}, namely,
\begin{eqnarray}\label{rge-functions}
\gamma_{\sigma}&=&\frac{1}{(4\pi)^3}\frac{Ng_1^2+g_2^2}{12}~,\nonumber\\
\beta_{g_1}&=&\frac{(N-8)g_1^3-12g_1^2g_2+g_1g_2^2}{12(4\pi)^3}~,\\ 
\beta_{g_2}&=&\frac{-4Ng_1^3+Ng_1^2g_2-3g_2^3}{4(4\pi)^3}~.\nonumber 
\end{eqnarray}

It should be noted that these functions were computed in the Minimal Subtraction (MS) renormalization scheme, and technically they should be adapted to our applications at hand, namely, the Coleman-
Weinberg procedure, as pointed out in~\cite{Ford:1991hw}. However, at the order we are interested here, that will not make any difference in our results, so it should not bother us any longer and we may put that matter aside. For a more detailed discussion, see for instance~\cite{Quinto:2014zaa}.

In order to solve (\ref{rge_seff}) and thus find the effective potential, we first observe that when $V_{eff}$ is calculated perturbatively the result can be organized as a power series in $L(\sigma)=\ln\frac{\sigma^2}{\mu^2}$, so we will assume the following {\it Ansatz}
\begin{equation}\label{Seff-ansatz}
 S_{eff}=A+BL+CL^2+DL^3+...
\end{equation}

\noindent where the coefficients are power series of the coupling constants $g_i$, that is,
\begin{equation}\label{coefficients1}
 A=\sum_{n=1}^\infty A_n, \quad \text{with} \quad \left\{\begin{array}{l}
A_1 = a_{11}g_1+a_{12}g_2  \\
A_2 = a_{21}g_1^2+a_{22}g_1g_2+a_{23}g_2^2 \\
A_3 = a_{31}g_1^3+a_{32}g_1^2g_2+a_{33}g_1g_2^2+a_{34}g_2^3 \\
\qquad\vdots
\end{array}\right.
\end{equation}

\noindent and similarly for the other coefficients, $B$, $C$, etc. 

The core idea behind the method is the observation that the coefficients in (\ref{Seff-ansatz}) are not all independent, since changes in $\mu$ must be compensated for by changes in the other 
parameters, according to the renormalization group. Let us then first reorganize the perturbative expansion (\ref{Seff-ansatz}) alternatively in the so-called leading-log series expansion. By simple power counting, we assemble the effective potential as follows
\begin{equation}\label{VLL}
 V_{eff}=\frac{\sigma^3}{6}\left(\sum_{n=0}^\infty C_n^{LL}g^{2n+1}L^n+\sum_{n=1}^{\infty}C_n^{NLL}g^{2n+3}L^n+\cdots+\delta\right),
\end{equation}
\noindent where $C_n^{LL}$ and $C_n^{NLL}$ are respectively the coefficients to the leading logarithms (LL) and next-to-leading logarithms (NLL) contributions, dots represent higher order contributions and 
$\delta$ is the counter-term defined by a renormalization condition. In the above expression, $g^{2n+1}$ denotes some combination of $g_1$ and $g_2$ at that order, such that $g^3$, for example, 
includes $g_1^3$, $g_1^2g_2$, $g_1g_2^2$ and $g_2^3$. 

To compute the leading-log contributions to the effective potential, we consider only the LL series,
\begin{equation}\label{LLpot}
 V_{eff}=\frac{\sigma^3}{6}\left(\sum_{n=0}^\infty C_n^{LL}g^{2n+1}L^n+\delta\right).
\end{equation}

\noindent In order to find the coefficients $C_n^{LL}$, we plug (\ref{Seff-ansatz}) in (\ref{rge_seff}) and consider each order in the expansion in $L$ to obtain the set of equations
\begin{eqnarray}
&& \left(\beta_1\frac{\partial}{\partial g_1} + \beta_2\frac{\partial}{\partial g_2} + 3\gamma_\sigma\right)A + 2(- 1 + \gamma_\sigma)B=0,  \qquad (\text{order}\quad L^0)\nonumber \\
&& \left(\beta_1\frac{\partial}{\partial g_1} + \beta_2\frac{\partial}{\partial g_2} + 3\gamma_\sigma\right)B + 2(- 1 + \gamma_\sigma)(2C)=0,  \qquad (\text{order}\quad L^1)\nonumber \\
&& \left(\beta_1\frac{\partial}{\partial g_1} + \beta_2\frac{\partial}{\partial g_2} + 3\gamma_\sigma\right)C + 2(- 1 + \gamma_\sigma)(3D)=0,  \qquad (\text{order}\quad L^2)\\
&& \qquad \vdots\nonumber
\end{eqnarray}
Now, each equation can also be expanded in powers of the coupling constants and thus we find:
\begin{eqnarray}
 && 2B_3=(\beta_1\frac{\partial}{\partial g_1}+\beta_2\frac{\partial}{\partial g_2}+3\gamma_\sigma)A_1 \qquad (\text{order}\quad g^3L^0),\nonumber \\
 && 4C_5=(\beta_1\frac{\partial}{\partial g_1}+\beta_2\frac{\partial}{\partial g_2}+3\gamma_\sigma)B_3 \qquad (\text{order}\quad g^5L^1),\nonumber \\
 && 6D_7=(\beta_1\frac{\partial}{\partial g_1}+\beta_2\frac{\partial}{\partial g_2}+3\gamma_\sigma)C_5 \qquad (\text{order}\quad g^7L^2),\\
&& \qquad \vdots \nonumber
\end{eqnarray}
where we have considered that $\gamma_\sigma \sim g^2$, $\beta_i \sim g^3$, $A_n \sim g^n$, $B_n \sim g^n$, etc. (cf. Eqs. (\ref{rge-functions}) and (\ref{coefficients1})).

The above set of equations allows us to identify the following recurrence relation for the LL coefficients
\begin{equation}\label{recurrence}
 C_{n+1}^{LL}=\left(\beta_1\frac{\partial}{\partial g_1}+\beta_2\frac{\partial}{\partial g_2}+3\gamma_\sigma\right)\frac{C_n^{LL}}{2(n+1)}.
\end{equation}

We are now able to compute the LL effective potential up to any order. In particular, it is important 
to note that the LL effective potential up to $g^3L$ order represents the full one-loop effective potential.

\subsection{The effective potential at one-loop order}\label{subseca}

We can now use (\ref{recurrence}) for $n=0$ to compute $C_{1}^{LL}$, with $C_{0}^{LL} = g_2$ being an input established from tree-level potential:
\begin{eqnarray}
C_{0}^{LL} & = & g_{2},\nonumber\\
C_{1}^{LL} & = &\left(\beta_1\frac{\partial}{\partial g_1}+\beta_2\frac{\partial}{\partial g_2}+3\gamma_\sigma\right)\frac{C_0^{LL}}{2}=-\frac{2g_1^3N-g_1^2g_2N+g_2^3}{256 \pi^3}.
\end{eqnarray}
 
The one-loop effective potential $V_{eff}$ is then given by
\begin{equation}\label{veff1}
V_{eff}= \frac{\sigma^3}{6} \left[g_2+\delta +\frac{1}{2} \left(\frac{g_2 \left(g_1^2 N +g_2^2\right)}{256 \pi^3}+\frac{-4 g_1^3 N +g_1^2 g_2 N -3 g_2^3}{256 \pi^3}\right)\ln \left(\frac{\sigma^2}{\mu^2}\right)\right].
\end{equation}

In order to fix the counter-term $\delta$, we use the Coleman-Weinberg renormalization condition,
\begin{equation}\label{cw1}
\frac{d^3V_{eff}}{d\sigma^3}\Big{|}_{|\sigma|=\mu}=g_2,
\end{equation}
\noindent where $\mu>0$ is the renormalization scale. Thus we find that the renormalized effective potential is
\begin{equation}\label{veff2}
V_{eff}= \frac{\sigma^3}{6}\left[g_2+
\frac{11\left(2 g_1^3 N-g_1^2 g_2 N+g_2^3\right)}{768 \pi^3}
-\frac{\left(2 g_1^3 N-g_1^2 g_2 N+g_2^3 \right)}{256 \pi^3}\ln\left(\frac{\sigma^2}{\mu^2}\right) \right].
\end{equation}

The classical potential is unbounded from below, but it is possible to have a metastable vacuum due to radiative corrections. Let us assume we have a local minimum and explore this possibility by imposing 
the renormalization scale to be around the (possible) local minimum of the effective potential. The conditions for its existence are given by 
\begin{subequations}\label{conditions}
\begin{align}
\frac{dV_{eff}}{d\sigma}\Big{|}_{|\sigma|=\mu} & = 0,\label{condition1} \\
\frac{d^2V_{eff}}{d\sigma^2}\Big{|}_{|\sigma|=\mu} & = m_{\sigma}^2>0,\label{condition2}
\end{align}
\end{subequations}
\noindent where $m_{\sigma}^2$ is the mass for the $\sigma$ field (possibly) generated by the radiative corrections.

\begin{figure}[h]
 \includegraphics[scale=0.6]{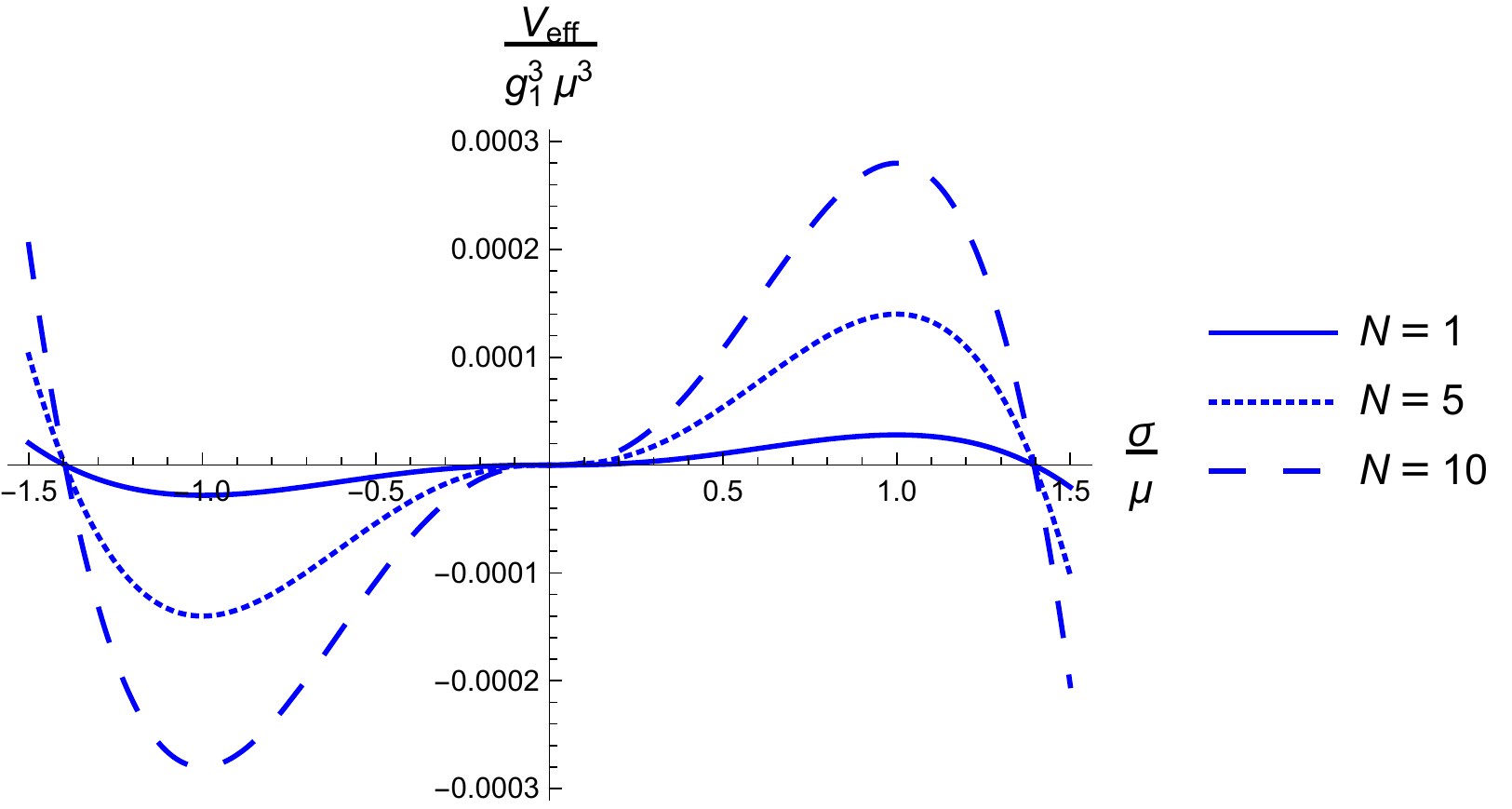}
 \caption{One-loop effective potential for different values of $N$. As large is $N$ as deeper is the valley of $V_{eff}$.}
 \label{fig1}
\end{figure}

Equation (\ref{condition1}) imposes that 
\begin{eqnarray}
g_2=-\frac{3}{256\pi^3}\left(2g_1^3 N-g_1^2 g_2 N+g_2^3\right),
\end{eqnarray}
\noindent and therefore the conditions (\ref{conditions}) are perturbatively satisfied for $\sigma=-\mu$ and $g_2\approx-\frac{3 g_1^3 N}{128\pi^3}$. Around the metastable vacuum, $V_{eff}$ can be written as
\begin{equation}\label{veff3}
V_{eff}= \frac{g_1^3 N \sigma^3}{2304\pi^3} \left[2-3 \ln \left(\frac{\sigma ^2}{\mu ^2}\right)\right],
\end{equation}
\noindent where the generated masses are given by
\begin{subequations}\label{masses}
\begin{align}
m_{\sigma}^2 & =\frac{d^2V_{eff}}{d\sigma^2}\Big{|}_{\sigma=-\mu}=\frac{g_1^3N}{128 \pi^3}\mu;\label{masssigma}\\
m_{\phi}^2 & =-g_1\langle\sigma\rangle=g_1\mu.\label{massphi}
\end{align}
\end{subequations}
\noindent We can see that both masses are positive, assuming $g_1>0$. The effective potential is plotted for different values of $N$ in figure \ref{fig1}.

The vacuum induced by radiative corrections is a local minimum and thus we have a metastable vacuum state and at some time it will decay to the real vacuum. However, the potential is unbounded from 
below, which means that there is no global minimum and therefore no stable solution to the potential with energy smaller than $\frac{-g_1^3 N \mu^3}{1152\pi^3}$.

Our results to the effective potential reveal three interesting phenomena. First, the model exhibit a dimensional transmutation, since the potential was initially described by two dimensionless parameters 
($g_1$ and $g_2$) and now it is described by a dimensionless parameter and a dimensionful one ($g_1$ and $\mu$, respectively). Second,  there is generation of mass to both fields in the $O(N)$-symmetric 
phase. Third, these phenomena are due to the appearance of a metastable vacuum.

The decay rate of the vacuum is in general computed through the Callan-Coleman formalism 
\cite{Callan:1977pt}, but this formalism can not be used in theories in which the symmetry breaking is due to radiative corrections, since it assumes a bounce solution to the classical potential. In order 
to compute such decays in theories in which spontaneous symmetry breaking is induced by radiative corrections, we apply a slightly changed form of the Callan-Coleman formalism developed by E. Weinberg \cite{Weinberg:1992ds}. 

In the case where there is no bounce solution (such as a potential unbounded from below) and the interactions are attractive, J. A. Gonz\'alez \textit{et al.} \cite{Gonzalez:2017hih} showed that 
there is no vacuum decay and the metastable vacuum is indeed the true vacuum. The authors carried out the analysis considering the Callan-Coleman formalism, but the results should be the same for Weinberg's formalism.

Physically, the tunnelling between false and true vacuum states occurs because when the system is in the false vacuum, quantum fluctuations creates bubbles of the true vacuum, continually. Now thinking 
about the tunnelling of the state as a phase transition, the bubble must be large enough to grow, i.e. a bubble with a sufficiently large radius to enclose the true vacuum solution.

However, the negative value of $g_2$ (assuming $g_1>0$) plays a central role in this analysis because, as the bubble grows, the repulsive interaction becomes more relevant. To see this, let us study the 
behaviour of the potential near the metastable vacuum, by Taylor expanding it to obtain,
\begin{eqnarray}
 V_{eff}&=&-\frac{g_1^3 N\mu^3}{1152 \pi^3}+
\frac{g_1^3 \mu  N (\sigma+\mu)^2}{256 \pi^3}-\frac{g_1^3 N (\sigma+\mu)^3}{256 \pi^3}+\mathcal{O}\left((\sigma+\mu)^4\right).
\end{eqnarray}

For small fluctuations around the local minimum of the effective potential $\sigma=-\mu$, this potential is similar to the discussed in \cite{Oliveira}, in this case the potential can simulate the 
dynamics of a long chain. In this way, when the bubble is large enough, the repulsive interaction becomes dominant and we observe the fracture of the chain. As expected, as $N$ grows, the metastable 
vacuum becomes more stable, once the $\phi$ fields interacts via an attractive interaction. This feature can be viewed graphically, because when $N$ is larger, the metastable vacuum is deeper, as showed in figure \ref{fig1}.

\subsection{The leading log effective potential}

Using the the recurrence relation (\ref{recurrence}), we can determine higher order corrections to the LL effective potential. The relevant observables of the theory around the metastable vacuum are 
sensitive up to $g^7L^3$ order, since the counter-term is determined up to $g^7$ order because renormalization condition (\ref{cw1}) . Therefore, in order to obtain the radiative generated masses 
it is enough to get only the first four terms in (\ref{recurrence}). Following the prescription described in the previous section, the renormalized LL effective potential up to $\mathcal{O}(g^7)$ is given by
\begin{eqnarray}\label{LL-Veff}
V_{eff} &=& \frac{\sigma^3}{6}\left[\tilde{A}+B\ln\left(\frac{\sigma^2}{\mu^2}\right)+C\ln^2\left(\frac{\sigma^2}{\mu^2}\right)+D\ln^3\left(\frac{\sigma^2}{\mu^2}\right)\right],
\end{eqnarray}
\noindent where
\begin{eqnarray}
\tilde{A} &=& g_2+\frac{g_1^7 N^3}{9437184 \pi^9}-\frac{7 g_1^7 N^2}{28311552 \pi^9}+\frac{5 g_1^7 N}{7077888 \pi^9}-\frac{5 g_1^6 g_2 N^3}{169869312 \pi^9}-\frac{109 g_1^6 g_2 N^2}{169869312 \pi^9}\nonumber\\
&&+\frac{73 g_1^6 g_2 N}{42467328 \pi^9}+\frac{g_1^5 g_2^2 N}{1179648 \pi^9}+\frac{g_1^5 N^2}{24576 \pi^6}-\frac{g_1^5 N}{12288 \pi^6}+\frac{5 g_1^4 g_2^3 N^2}{56623104 \pi^9}-\frac{g_1^4 g_2^3 N}{18874368 \pi^9}\nonumber\\
&&-\frac{g_1^4 g_2 N^2}{73728 \pi^6}-\frac{7 g_1^4 g_2 N}{73728 \pi^6}+\frac{11 g_1^3 g_2^4 N}{28311552 \pi^9}+\frac{11 g_1^3 N}{2304 \pi^3}-\frac{49 g_1^2 g_2^5 N}{169869312 \pi^9}+\frac{g_1^2 g_2^3 N}{36864 \pi^6}\nonumber\\
&&-\frac{11 g_1^2 g_2 N}{4608 \pi^3}+\frac{7 g_2^7}{18874368 \pi^9}-\frac{g_2^5}{24576 \pi^6}+\frac{11 g_2^3}{4608 \pi^3}\nonumber\\
B&=& \frac{g_1^2 N (g_2-2 g_1)-g_2^3}{1536 \pi^3},\nonumber\\
C&=& \frac{-3 g_1^5 (N-2) N+g_1^4 g_2 N (N+7)-2 g_1^2 g_2^3 N+3 g_2^5}{589824 \pi^6},\nonumber\\
D&=& -\frac{g_1^7 N^3}{75497472 \pi^9}+\frac{7 g_1^7 N^2}{226492416 \pi^9}-\frac{5 g_1^7 N}{56623104 \pi^9}+\frac{5 g_1^6 g_2 N^3}{1358954496 \pi^9}+\frac{109 g_1^6 g_2 N^2}{1358954496 \pi^9}\nonumber\\
&&-\frac{73 g_1^6 g_2 N}{339738624 \pi^9}-\frac{g_1^5 g_2^2 N}{9437184 \pi^9}-\frac{5 g_1^4 g_2^3 N^2}{452984832 \pi^9}+\frac{g_1^4 g_2^3 N}{150994944 \pi^9}-\frac{11 g_1^3 g_2^4 N}{226492416 \pi^9}\nonumber\\
&&+\frac{49 g_1^2 g_2^5 N}{1358954496 \pi^9}-\frac{7 g_2^7}{150994944 \pi^9}.\nonumber
\end{eqnarray}

Just as in the one-loop case, the conditions (\ref{conditions}) are perturbatively satisfied for $\sigma=-\mu$, but the coupling constant $g_2$ receives corrections up to $\mathcal{O}(g_1^7)$ given by
\begin{eqnarray}
g_2= -\frac{3 g_1^3 N}{128 \pi^3}
-\frac{g_1^5 N\left(17 N-16\right)}{32768 \pi^6}
-\frac{g_1^7 N\left(651 N^2+464 N+320\right)}{75497472 \pi^9}.
\end{eqnarray}

Therefore, the LL effective potential is
\begin{eqnarray}\label{veff3l}
V_{eff}&=&\frac{g_1^3 N \sigma ^3}{2304 \pi^3}\Big[2+\frac{g_1^4 N (17 N-16)+768 \pi^3 g_1^2 N}{32768 \pi^6}
-\left(3+\frac{3 g_1^2 N \left(g_1^2 (17 N-16)+768 \pi^3\right)}{65536 \pi^6}\right) \ln\left(\frac{\sigma^2}{\mu^2}\right)\nonumber\\
&&-\frac{3 \left(g_1^4 N (N+7)+128 \pi^3 g_1^2 (N-2)\right)}{32768 \pi^6} \ln^2\left(\frac{\sigma^2}{\mu^2}\right)
-\frac{g_1^4 ((7-3 N) N-20)}{98304 \pi^6} \ln^3\left(\frac{\sigma^2}{\mu^2}\right)\Big].
\end{eqnarray}

\begin{figure}[h]
 \includegraphics[scale=0.6]{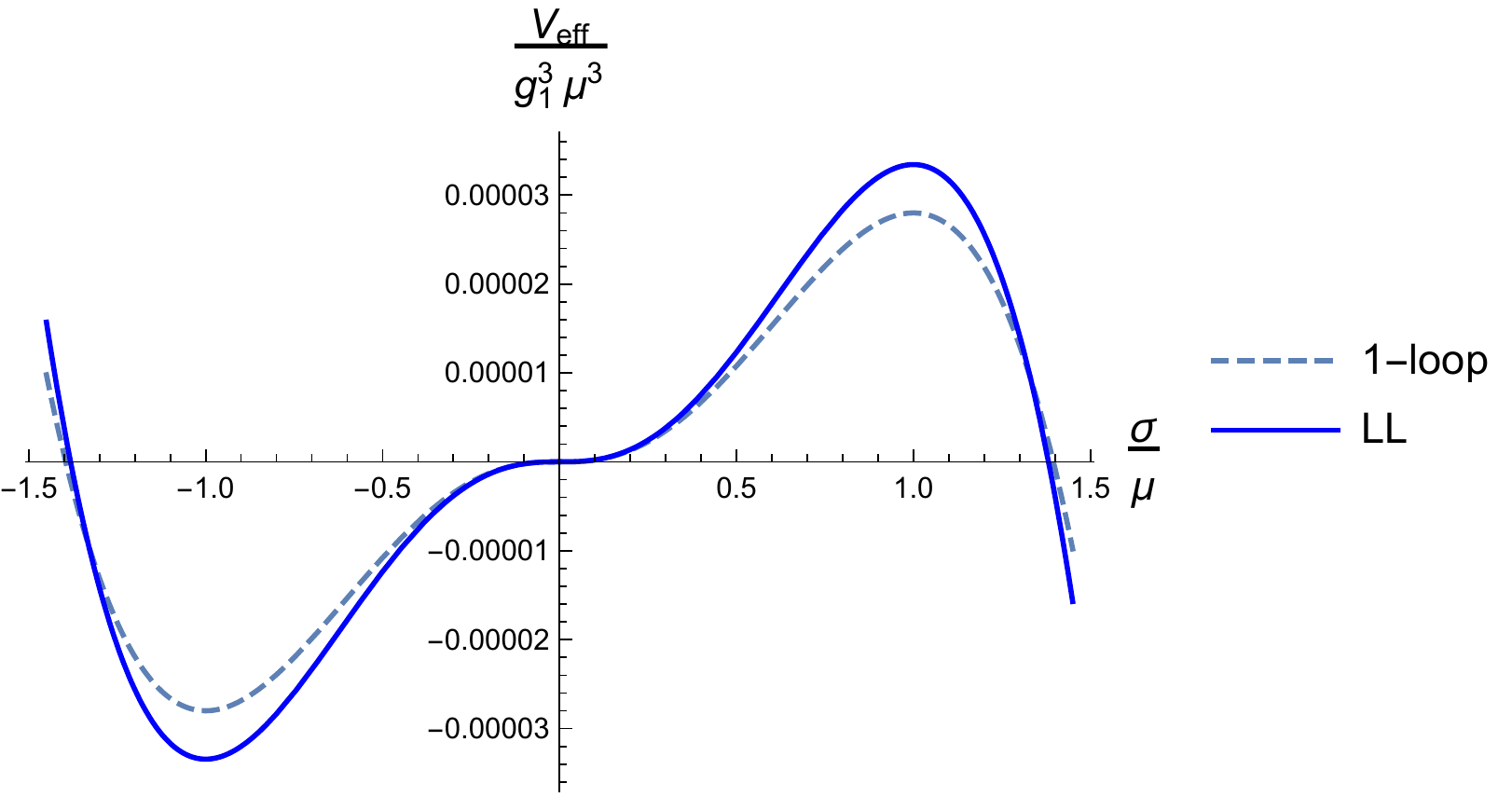}
 \caption{Comparison between one-loop Eq.(\ref{veff3}) and LL Eq.(\ref{veff3l}) effective potentials for $N=10^4$ and $g_1=0.2$. Leading Log corrections become relevant for large $N$.}
 \label{fig2}
\end{figure}

The fields acquire mass induced by radiative corrections given by
 \begin{eqnarray}
  &&m_\sigma^2=\frac{d^2V_{eff}}{d\sigma^2}\Bigr|_{\sigma=-\mu}=\frac{g_1^3N\mu}{128\pi^3}\left[1+\frac{g_1^2(13N-8)}{768\pi^3}+\frac{g_1^4N(59N+8)}{196608\pi^6}\right],
 \end{eqnarray}

\noindent where $m_{\phi}^2$ is the same as (\ref{massphi}). The LL corrections to $m_\sigma^2$ become larger as $N$ grows. For instance, if we have $g_1\sim 0.2$ and $N\sim 10^3$, the corrections to the 
one-loop mass is of order of $2\%$, and for $N\sim 10^4$ the corrections is about $27\%$. Therefore, the LL corrections becomes very relevant in the large $N$ limit of the effective potential. In the 
figure \ref{fig2} we plot the comparison between one-loop (\ref{veff3}) and LL (\ref{veff3l}) effective potentials for $N=10^4$ and $g_1=0.2$.   

\subsection{The large $N$ limit}

One interesting case to explore general $O(N)$ models is the large $N$ expansion~\cite{tHooft:2002ufq}. The large $N$ expansion is an alternative way to organize the 
perturbative series and has important phenomenological applications, such as in QCD~\cite{tHooft:2003lzk} and condensed matter phenomena, e.g., through non-linear sigma models
~\cite{Arefeva:1979bd,Arefeva:1978fj,Arefeva:1980ms,Rosenstein:1989sg}. To discuss some features of a large $N$ limit in the six dimensional cubic theory, let us redefine the coupling constants in (\ref{o(n)-lagrangian}) as $g_1\rightarrow g_1/\sqrt{N}$ and $g_2\rightarrow g_2/\sqrt{N}$, 
\begin{eqnarray}\label{o(n)-lagrangian-1}
 \mathcal{L}=\frac{1}{2}(\partial_\mu\phi^i)^2+\frac{1}{2}(\partial_\mu\sigma)^2+\frac{g_1}{2\sqrt{N}}(\sigma\phi^i\phi^i)+\frac{g_2}{6\sqrt{N}}\sigma^3, \qquad (i=1,2,\cdots, N).
\end{eqnarray}

Through this redefinition, the RGE functions (\ref{rge-functions}) become
\begin{eqnarray}\label{rge-functions-1}
\gamma_{\sigma}&=&\frac{1}{(4\pi)^3}\frac{g_1^2}{12}~,\nonumber\\
\beta_{g_1}&=&\frac{g_1^3}{12\sqrt{N}(4\pi)^3}~,\\ 
\beta_{g_2}&=&\frac{-4g_1^3+g_1^2g_2}{4\sqrt{N}(4\pi)^3}~.\nonumber 
\end{eqnarray}

The effective potential in the large $N$ expansion have the general structure:
\begin{eqnarray}
V_{eff} &=& \frac{\sigma^3}{6\sqrt{N}}\left(\sum_{i=0}^{\infty}C_iL^i+\frac{1}{N}\sum_{i=0}^{\infty}C_iL^i+\cdots\right),
\end{eqnarray}
from which we can see that while the LL expansion (\ref{VLL}) is based on a relation between powers of $L$ and the coupling constants, in the large $N$ expansion we can have a complete series of $L$ at same leading order of $N$. 

In our case, to obtain a large $N$ effective potential up to some order of $L$ (or coupling constants), we can just use the LL effective potential presented in the previous section. In fact, the 
leading power terms of $N$ can be taken from (\ref{LL-Veff}) and then we can apply the redefinition of the coupling constants $g_1\rightarrow g_1/\sqrt{N}$ and $g_2\rightarrow g_2/\sqrt{N}$ to obtain
\begin{eqnarray}
V_{eff} &=& \frac{\sigma^3}{6\sqrt{N}}\left[\tilde{A}+B\ln\left(\frac{\sigma^2}{\mu^2}\right)+C\ln^2\left(\frac{\sigma^2}{\mu^2}\right)+D\ln^3\left(\frac{\sigma^2}{\mu^2}\right)\right],
\end{eqnarray}
\noindent with
\begin{eqnarray}
\tilde{A} &=& g_2+\frac{g_1^7}{9437184 \pi^9}-\frac{5 g_1^6 g_2}{169869312 \pi^9}+\frac{g_1^5}{24576 \pi^6}-\frac{g_1^4 g_2}{73728 \pi^6}+\frac{11 g_1^3}{2304 \pi^3}-\frac{11 g_1^2 g_2}{4608 \pi^3},\nonumber\\
B&=& \frac{g_1^2(g_2-2 g_1)}{1536 \pi^3},\nonumber\\
C&=& \frac{-3 g_1^5+g_1^4 g_2}{589824 \pi^6},\nonumber\\
D&=& -\frac{g_1^7}{75497472 \pi^9}+\frac{5 g_1^6 g_2}{1358954496 \pi^9},\nonumber
\end{eqnarray}

\noindent where the counter-term $\delta$ was fixed by the condition
\begin{equation}\label{cw1-N} 
\frac{d^3V_{eff}}{d\sigma^3}\Big{|}_{|\sigma|=\mu}=\frac{g_2}{\sqrt{N}}. 
\end{equation}

The CW effective potential is then given by
\begin{eqnarray}\label{veff3l-N}
V_{eff}&=&\frac{g_1^3\sigma ^3}{2304 \pi^3\sqrt{N}}\Big[2+\frac{g_1^2(17g_1^2+768 \pi^3)}{32768 \pi^6}
-\left(3+\frac{3 g_1^2\left(17g_1^2+768 \pi^3\right)}{65536 \pi^6}\right) \ln\left(\frac{\sigma^2}{\mu^2}\right)\nonumber\\
&&-\frac{3g_1^2 \left(g_1^2 +128 \pi^3\right)}{32768 \pi^6} \ln^2\left(\frac{\sigma^2}{\mu^2}\right)
+\frac{g_1^4}{32768 \pi^6} \ln^3\left(\frac{\sigma^2}{\mu^2}\right)\Big],
\end{eqnarray}

\noindent where the conditions (\ref{conditions}) were satisfied for $\sigma=-\mu$, with the following relation between the coupling constants
\begin{eqnarray}
g_2= -\frac{3 g_1^3}{128 \pi^3}\left[1
+\frac{17 g_1^5}{768 \pi^6}+\frac{217g_1^7}{589824 \pi^9}\right].
\end{eqnarray}

One interesting feature of large $N$ expansion is that at same order of $N$ we have contributions of different orders in coupling constants, as we can see from (\ref{veff3l-N}).

In the large $N$ limit, the mass of the $\sigma$ field becomes
 \begin{eqnarray}
  &&m_\sigma^2=\frac{d^2V_{eff}}{d\sigma^2}\Bigr|_{\sigma=-\mu}=\frac{g_1^3\mu}{128\pi^3\sqrt{N}}\left[1+\frac{13g_1^2}{768\pi^3}+\frac{59g_1^4}{196608\pi^6}\right].
 \end{eqnarray}

It is important to note that higher order powers of $L$ will not contribute to the generated mass because the counter-term $\delta$ receives corrections only up to the $L^3$ order, due to renormalization condition (\ref{cw1-N}).    

\section{Conclusions}

In this work, we studied the possibility of a spontaneous generation of mass, induced by radiative corrections via Coleman-Weinberg mechanism, in a model consisting of $N$ scalar fields $\phi^i$ 
transforming in the fundamental representation of $O(N)$ coupled to an additional scalar field $\sigma$ via cubic interactions, defined in a six dimensional spacetime. We computed the improved 
effective potential and use it to discuss the vacuum structure of the model. This model has a potential unbounded from below, but it is nevertheless possible that radiative corrections might 
generate a stable false vacuum, as discussed in \cite{Gonzalez:2017hih}. Our results indicate that the Coleman-Weinberg mechanism does indeed provide a metastable vacuum and a generation of mass in the model presented here.  

\vspace{.5cm}
{\bf Acknowledgements.} 
This work was partially supported by Conselho Nacional de Desenvolvimento Cient\'{\i}fico e 
Tecnol\'{o}gico (CNPq). A.C.L. has been partially supported by the CNPq project 402096/2016-9. The 
work by H.S. has been partially supported by the program PIBIC/CNPq through the project 402096/2016-9.

\appendix
\section{The (metastable) vacuum structure}

In the main text, since we are only interested in the mass generation, we have assumed that $\langle\phi_i\rangle=0$ and thus $\sigma$ is the only degree of freedom in the effective potential. In 
this appendix we discuss the vacuum structure of the model and show that it has two phases, only one with mass generation.

We start shifting the quantum fields in (\ref{o(n)-lagrangian}) by the corresponding classical field backgrounds 
\begin{equation}\label{shift}
\phi^N  \longrightarrow (\phi^N-\varphi_c)\qquad \text{and} \qquad \sigma \longrightarrow (\sigma-\sigma_c).
\end{equation}
As the careful reader will notice, in our work we have implicitly considered a shift to the left for $
\sigma$ and thus a minimum of the potential in a negative value of the field. In this appendix, however, such choice would be inconvenient since we are dealing with propagators, so here we have made the choice above instead.

In terms of the redefined fields (\ref{shift}), the Lagrangian is
\begin{eqnarray}\label{o(n-1)-lagrangian}
 \mathcal{L}&=&\frac{1}{2}(\partial_\mu\phi^i)^2+\frac{1}{2}(\partial_\mu\sigma)^2+\frac{g_1}{2}(\sigma\phi^i\phi^i)+\frac{g_2}{6}\sigma^3-\frac{g_1\sigma_c}{2}(\phi^i)^2\nonumber\\
&& +\frac{1}{2}\left(3g_2\sigma_c^2+g_1\varphi_c^2\right)\sigma
 +g_1\sigma_c\varphi_c\phi^N
 -\frac{3g_2\sigma_c}{2}\sigma^2-g_1\varphi_c\sigma\phi^N, \qquad (i=1,2,\cdots, N).
\end{eqnarray}

Due the presence of a term that mixes $\sigma$ and $\phi^N$, the quadratic part of the Lagrangian can be written as
\begin{eqnarray}
\mathcal{L}=-\frac{1}{2}\left(\sigma,\phi^N  \right)\left(\begin{array}{c c }
\Box+3g_2\sigma_c^2 & g_1\sigma_c\varphi_c ‚Ä¶ \\
g_1\sigma_c\varphi_c & \Box+g_1\sigma_c^2
\end{array} \right)
\left(\begin{array}{c}
\sigma \\
\phi^N 
\end{array} \right)-\frac{1}{2}\phi^j\left(\Box+g_1\sigma_c^2 \right)\phi^j+\cdots, 
\end{eqnarray}
\noindent where $j=1,2,\cdots, (N-1)$.

The propagators of the model in the momenta space can be cast as
\begin{subequations}\label{propagators}
\begin{align}
\langle T\phi^j(p)\phi^j(-p)\rangle & =\frac{i}{p^2-g_1\sigma_c} \qquad (j=1,2,\cdots, N-1)~,\label{prop_phi}\\
\langle T\phi^N(p)\phi^N(-p)\rangle & =\frac{i(p^2-g_1\sigma_c) }{p^2(p^2-3g_2\sigma_c-g_1\sigma_c)-g_1^2 \sigma_c^2 \varphi_c^2+3g_1g_2 \sigma_c^2}~,\label{prop_phiN}\\
\langle T\sigma(p)\sigma(-p)\rangle & =\frac{i(p^2-3g_2\sigma_c)}{p^2(p^2-3g_2\sigma_c-g_1\sigma_c)-g_1^2 \sigma_c^2 \varphi_c^2+3g_1g_2 \sigma_c^2}~,\label{prop_sigma} \\
\langle T\phi^N(p)\sigma(-p)\rangle & =\frac{i g_1\sigma_c\varphi_c }{p^2(p^2-3g_2\sigma_c-g_1\sigma_c)-g_1^2 \sigma_c^2 \varphi_c^2+3g_1g_2 \sigma_c^2}~.\label{prop_phi_sigma}
\end{align}
\end{subequations}
\noindent It is easy to see that for $\sigma_c=0$ the above propagators reduce to propagators of massless fields and $\langle T\phi^N(p)\sigma(-p)\rangle$ vanishes. Taking $\varphi_c=0$, such propagators reduce to
\begin{subequations}\label{propagators_reduced}
\begin{align}
\langle T\phi^j(p)\phi^j(-p)\rangle & =\frac{i}{p^2-g_1\sigma_c} \qquad (i=1,2,\cdots, N)~,\label{prop_phi_b}\\
\langle T\sigma(p)\sigma(-p)\rangle & =\frac{i }{p^2-3g_2\sigma_c}~,\label{prop_sigma_b} \\
\langle T\phi^N(p)\sigma(-p)\rangle & =0~.\label{prop_phi_sigma_b}
\end{align}
\end{subequations}

At tree level, from (\ref{o(n-1)-lagrangian}), we see that the tadpole equations for $\sigma$ and $\phi^N$ are given by
\begin{subequations}\label{gap_tree}
\begin{align}
\langle \sigma(0)\rangle & = \frac{i}{2}\left(3g_2\sigma_c^2+g_1\varphi_c^2 \right)=0~,\label{tadpole_sigma_tree} \\
\langle \phi^N(0)\rangle & = -ig_1\sigma_c\varphi_c=0~,\label{tadpole_phi_tree}
\end{align}
\end{subequations}
\noindent which possess only $\sigma_c=\varphi_c=0$ as solution. 

But the situation is different at one-loop level. In fact, due to the mixing $\langle T\phi^N(p)\sigma(-p)\rangle$, the tadpole equation for $\phi^N$ is given by
\begin{eqnarray}\label{gap_1loop_phi}
\langle \phi^N(0)\rangle & = & -ig_1\sigma_c\varphi_c\Big\{1
-\frac{i}{2} \int{\frac{d^6k}{(2\pi)^6}}\frac{1}{k^2(k^2-3g_2\sigma_c-g_1\sigma_c)-g_1^2 \sigma_c^2 \varphi_c^2+3g_1g_2 \sigma_c^2}\Big\}=0~,
\end{eqnarray}
\noindent where it is easy to see that the expression inside brackets is non-vanishing, so the condition $\sigma_c\varphi_c=0$ still holds. 

Choosing $\varphi_c=0$, we have the following $\sigma$ tadpole equation
\begin{eqnarray}\label{gap_1loop_sigma}
\langle \sigma(0)\rangle & = & \frac{3i}{2}g_2\sigma_c^2
-\frac{g_2}{2}\int{\frac{d^6k}{(2\pi)^6}}\frac{1}{k^2-3g_2\sigma_c}
-\frac{Ng_1}{2}\int{\frac{d^6k}{(2\pi)^6}}\frac{1}{k^2-g_1\sigma_c}=0.
\end{eqnarray}

The integral over $k$ is given by
\begin{eqnarray}
\int{\frac{d^{(6-2\epsilon)}k~\tilde{\mu}^{2\epsilon}}{(2\pi)^6}}\frac{-i}{k^2-m^2}=
\frac{m^4}{128 \pi^3 \epsilon }-\frac{m^4}{256 \pi^3}\left[2 \ln \left(\frac{m^2}{\tilde{\mu}^2}\right)+C\right]+\mathcal{O}\left(\epsilon\right)~,
\end{eqnarray}
\noindent where $C=2 \gamma -3+2 \log (\pi )$ and $\gamma$ is the Euler-Mascheroni constant.

Therefore, (\ref{gap_1loop_sigma}) can be cast as
\begin{eqnarray}\label{gap_1loop_sigma2}
\langle \sigma(0)\rangle & = & \frac{i}{2}\sigma_c^2\left[3 g_2-\frac{9g_2^3+Ng_1^3}{256\pi^3\epsilon}
+\frac{9g_2^3+Ng_1^3}{512\pi^3\epsilon}\ln\left(\frac{\sigma_c^2}{\mu^2}\right)+\tilde{C}(g_1,g_2)\right]=0,
\end{eqnarray}

\noindent where $\tilde{C}(g_1,g_2)$ is a constant function of $g_1$ and $g_2$, and $\mu=\tilde{\mu}^2$ is a scale with the same dimension as $\sigma_c$.

The pole ($\epsilon=0$) can be MS removed by the introduction of a linear contra-term (see for instance \cite{Collins:1984xc}). It is easy to see that one possible solution to the tadpole equation 
(\ref{gap_1loop_sigma2}) is $\sigma_c=0$. But $\sigma_c=0$ corresponds to an inflection point of the effective potential, as we see from the figure (\ref{fig1}). The solution $\sigma_c=+\mu~f(g_1,g_2)\neq0$, with $f(g_1,g_2)$ being an exponential function of $g_1$ and $g_2$, corresponds 
to the local minimum of $V_{eff}$, while $\sigma_c=-\mu~f(g_1,g_2)$ to the local maximum. The difference in the value of $\sigma_c$ found here and in the section \ref{subseca} is due to the 
difference on renormalization schemes (CW renormalization scheme for $V_{eff}$ and MS for the tadpole equation) and has no physical significance.  

Like the non-linear sigma models \cite{Arefeva:1979bd,Arefeva:1978fj,Arefeva:1980ms,Rosenstein:1989sg}, the present theory exhibits two 
phases. In the $O(N)$ symmetric phase, $\varphi_c^N=0$ and $\sigma_c\neq0$, the model presents spontaneous generation of mass (due to generation of a metastable vacuum as discussed in this 
article). In the other phase, $\varphi_c^N\neq0$ and $\sigma_c=0$, the $O(N)$ symmetry is spontaneously broken to $O(N-1)$ and there is no spontaneous mass generation.

\end{document}